\documentclass[letterpaper,twocolumn,prl,superscriptaddress]{revtex4}
\usepackage{bm}
\usepackage{epsfig}
 
\begin{document}

\title{Role of inertia in two-dimensional deformation and breakup of a
droplet}

\author{A. J. Wagner}
\email{Alexander.Wagner@ndsu.nodak.edu}
\affiliation{Dept. of Physics, NDSU, Fargo 58105, ND, USA}
\affiliation{School of Physics, University of Edinburgh, JCMB Kings
Buildings, Mayfield Road, Edinburgh EH9 3JZ, Scotland}
\author{L. M. Wilson}
\affiliation{School of Physics, University of Edinburgh, JCMB Kings
Buildings, Mayfield Road, Edinburgh EH9 3JZ, Scotland}
\author{M. E. Cates}
\affiliation{School of Physics, University of Edinburgh, JCMB Kings
Buildings, Mayfield Road, Edinburgh EH9 3JZ, Scotland}

\begin{abstract}
We investigate by Lattice Boltzmann methods the effect of inertia on
the deformation and break-up of a two-dimensional fluid droplet
surrounded by fluid of equal viscosity (in a confined geometry) whose
shear rate is increased very slowly. We give evidence that in two
dimensions inertia is {\em necessary} for break-up, so that at zero
Reynolds number the droplet deforms indefinitely without breaking. We
identify two different routes to breakup via two-lobed and three-lobed
structures respectively, and give evidence for a sharp transition
between these routes as parameters are varied. 
\end{abstract}
\maketitle

The role of inertia in the deformation and breakup of a droplet in a
shear flow remains an open problem in fluid mechanics. In three
dimensions, a droplet in a steady shear field will break at a critical
capillary number Ca $\simeq 0.5$ even without inertia
\cite{taylor}. Crudely speaking, this happens when it is deformed
enough to undergo the Rayleigh instability, which is the peristaltic
instability of a long tube of one static fluid in another, driven by
interfacial tension. Ca measures the relative importance of viscous to
interfacial forces; the effect of inertia, whose importance is
quantified by the Reynolds number Re, is essentially
perturbative. Thus, although situations arise where a droplet that
would be stable at zero Re is unstable at a finite Re, this really
amounts to an inertia-induced shift of the critical Ca. The
contribution of inertia was studied in recent numerical work
\cite{renardy1,renardy2,xi,li}. Much of this work uses a flow protocol
based on sudden onset of shearing at different flow rates, rather than
a gradual ramping up of the rate; this saves much time numerically but
describes a distinct physical situation from the slow ramp studied
here; both are of interest experimentally.

The situation is somewhat different in two dimensions. Although not
realizable directly in laboratory experiments, this case is more
accessible by simulation \cite{halliday,sheth}, and is interesting
because the nonlinear physics involved in breakup may not be the same
as in 3D.  In 2D, the Rayleigh instability is effectively switched
off, because periodic variations in the width of the droplet (at fixed
volume) always increases the amount of interface. In the present work
we study numerically in 2D the breakup of a droplet, and find quite
distinctive nonlinear physics, in which inertia plays an essential
part. The relatively moderate demands of 2D simulations allows us to
establish the scenario relatively clearly, even for the case of an
`infinitesimal' ramp, which we here approach numerically by a novel
recursive updating scheme for the flow rate.

{\em Method:} We use the lattice Boltzmann method to simulate a
(symmetric) binary fluid deep within the two-phase region. The phase
separation of the order parameter $\phi$ is induced by a chemical
potential $\mu=A\phi+B\phi^3-\kappa \nabla^2 \phi$ derived from a
$\phi^4$ Landau free energy \cite{ludwig}. The evolution equation for
the order parameter is
\begin{equation}
\partial_t \phi + {\bf u}.\nabla \phi = M\nabla^2\mu
\end{equation}
with $M$ a mobility; this is coupled with a Navier-Stokes equation
\begin{equation}
\rho (\partial_t {\bf u}+{\bf u}.\nabla{\bf u})
=-\nabla p +\phi\nabla\mu +\eta \nabla^2 {\bf u}
\end{equation}
where ${\bf u}$ is the fluid velocity, $\rho$ its density and $\eta$
is the viscosity. The interfacial tension is included in this model
through the gradient terms in the chemical potential and it is given
by $\sigma=\frac{2A}{3B}\sqrt{-2\kappa A}$. Note that topological
reconnections of the fluid-fluid interface are handled implicitly by
the order parameter diffusion and no singularity is encountered at
breakup. Care was taken with parameter selection to ensure that the
droplet behaviour is dominated by hydrodynamic and capillary forces,
except very close to the breakup point, where both diffusion, and an
interaction arising from the finite breadth of the interfaces (a few
lattice spacings), kick in.
 
In our simulations, we first set up a droplet of fluid in a
surrounding immiscible phase of the same viscosity. To impose a shear
flow with strain rate $\dot\gamma$, we introduce moving walls at the
top and bottom edges of the simulation cell with opposite velocities
$\pm L\dot\gamma/2$ where $L$ is the cell width. A droplet in such a
shear profile will deform with the flow until and unless the viscous
stress is balanced by interfacial tension. The ratio of these forces
is expressed in the capillary number
\begin{equation}
\mbox{Ca} = \frac{\eta \dot{\gamma} R_0}{\sigma}
\end{equation}
where $R_0$ is the radius of the undeformed droplet; the Reynolds
number measures the ratio of inertial and viscous forces:
\begin{equation}
\mbox{Re} = \frac{\rho \dot{\gamma} R_0^2}{\eta}
\end{equation}

We quantify the deformation of a droplet by calculating the following
moments of the order parameter field
\begin{eqnarray}
b &=& \sum_{\bf x} \theta(\phi) \phi\\
c_\alpha &=& \sum_{\bf x} \theta(\phi) x_\alpha\\
d_{\alpha\beta} &=& \sum_{\bf x} \theta(\phi) \phi (x_\alpha-c_\alpha)
(x_\beta-c_\beta) 
\end{eqnarray}
Here $b$ gives the 2D droplet's area and ${\bf c}/b$ the position
vector of its centre. The eigenvalues of $d_{\alpha\beta}/b$ give us
two length scales $l_1\ge l_2$ corresponding to the two axes of the
droplet. The Taylor deformation is defined as
\begin{equation}
D = \frac{l_1-l_2}{l_1+l_2}.
\end{equation}
Note that very deformed droplets have $D$ close to (but still less
than) unity.

To study the breakup behavior of droplets and find the critical
capillary number (beyond which no steady state for a single droplet
stably exists) we need to find the stationary shapes of droplets at
various Ca and Re. We ramp up the shear rate in a way that finds all
stable droplet shapes before the breakup. To do this efficiently we
use an algorithm that increases the shear rate $\dot{\gamma}$ stepwise
and then waits for the droplet shape to reach steady-state before
calculating Ca and Re. If there is breakup, or if the resulting
deformation increment is too large (exceeding a threshold $\Delta D_c
\simeq 0.05$) a smaller increase $\Delta\dot\gamma$ in the shear rate
is tried.  This process is iterated until either equilibrium is
achieved, or $\Delta\dot\gamma$ falls below a pre-set limit chosen
very close to zero ($\Delta\dot\gamma_c\simeq 10^{-7}$).  In the
latter case, the last value of $\dot\gamma$ for which equilibrium {\em
was} reached is identified as the final stable shear rate; this fixes
the value of Ca and Re at breakup for the chosen run. By varying
simulation parameters (such as viscosity) an ensemble of curves of Re
vs Ca was generated. Each termininated in a breakup point.

The simulations reported below have lattices bounded by moving walls
separated by 120 lattice sites, with periodic boundary conditions
along the flow direction (in which the lattice length is 400). Other
parameters common to all runs were, in lattice units \cite{kendon},
$-A=B = 0.06; \kappa = 0.039$, from which the interfacial tension is
derived as $0.046$; and $M = 3.0$. Ten viscosity values were selected
in the range $0.23-0.88$, again in lattice units \cite{kendon}. These
parameter values were chosen to minimise lattice anisotropy, spurious
diffusive currents, and other well-documented but fairly controllable
artefacts of the lattice Boltzmann method \cite{kendon,foot}. The
initial droplet radius in each run was 21 lattice units. However, it
was found that the increased curvature of a sheared droplet could lead
to slight changes in its size as a result of increased Laplace
pressure causing a shift in solubility. This effect was minimized by
using the true size of the droplet when calculating $R_0$ for the
purposes of Re and Ca (see below).

The ratio $L/R_0 \simeq 5.7$ is not so large as to approach the
behaviour of a droplet being sheared in an infinite medium. Wall
proximity effects are present in our results (as they are in many
experiments), but we do not study these systematically here. In
addition, the slight change in $R_0$ with shear rate mentioned above
means that the wall effects vary slightly during each run. By varying
lattice size and shape we checked that there was no undue influence on
the results. The same applies to the effect caused by proximity of a
droplet to its periodic image along the flow direction.

Although it is hard to quantify precisely all systematic errors in
lattice Boltzmann \cite{kendon} we believe the quantitative results
reported below are generally good to within ten percent and mostly
rather better. Our main conclusions, however, are qualitative in
nature.

\begin{figure}
\resizebox{\columnwidth}{!}{\rotatebox{90}{\includegraphics{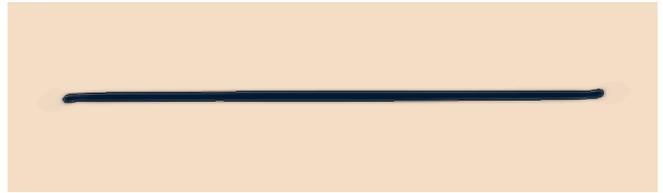}}}
\caption{For zero Reynolds number a two dimensional droplet does not
break up. It extends until the width of the droplet is of the order of
the interface width. The picture shows a droplet with Ca $= 1.1$; the
upper wall is moving rightward, the lower, leftward.}
\label{fignoRE}
\end{figure}

{\em Zero Reynolds number results:} First we report a simulation for
zero Reynolds number. While lattice Boltzmann fully treats inertial
effects, a stationary zero Reynolds number flow can also be simulated
(essentially by switching off the term in the algorithm that handles
fluid convection). In contrast to previous suggestions
\cite{halliday,wagner}, we found that a two-dimensional droplet in a
shear flow at zero Re {\em does not break up} by any fluid-mechanical
process. Instead it continues to deform (Figure \ref{fignoRE}) until
the width of the droplet is comparable to the interface width, at
which point dissolution-assisted breakup occurs. This corresponds to a
molecular rather than a hydrodynamic breakup mechanism. A similarly
elongated droplet in 3D would certainly not survive, due to the
Rayleigh instability.

\begin{figure}
\resizebox{\columnwidth}{!}{\includegraphics{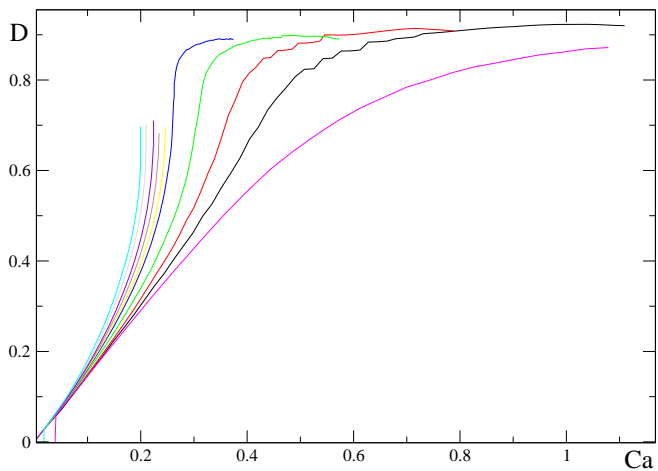}}\\
\caption{Taylor deformation vs. capillary number under slow ramping of
shear rates for (top to bottom) nine runs of viscosities in the range
$0.181-0.877$ (lattice units) and the zero Re limit. Values of the
parameter $I =$ Re/Ca are $I = 18, 13, 11, 9.1, 7.3, 6.3, 3.7, 2.2,
1.3, 0$ (top to bottom).}
\label{figD}
\end{figure}

{\em Finite Reynolds number results:} Both the Reynolds number and the
capillary number depend linearly on the shear rate. We can, however,
form the dimensionless number $I =$ Re/Ca which will be almost
constant for each run (deviations arising only from the small changes
in droplet size $R_0$; see above).  The nominal $I$ values for the
nine different runs simulated are identified in the caption to Figure
\ref{figD}, where we show deformation curves in each case. At modest
Ca values the curves with a higher $I$ (more inertia) show a larger
deformation. A likely explanation for this is the Bernoulli
effect. This will create regions of lower pressure at the edges of the
droplet, tending to stretch it into the flow so that it experiences a
stronger flow than applies for a droplet at zero Re.

The simulation corresponding to the lowest curve is the zero Reynolds
number case and here the droplet does not break, as previously
discussed. All other curves end at the last stable configuration for
the droplet that could be found using our ramp method. It can be
difficult to establish the stability of the very slowly evolving,
near-critical droplets that arise close to this last stable
configuration (particularly if they are very deformed) so we have
taken care to run the simulations for tens of thousands of timesteps
per iteration in these parts of the curves.

With increasing $I$ we find that there is a trend towards steeper
deformation curves, and at a critical $I_c$ between $6.3$ and $7.3$ we
observe a singularity in which the deformation curve appears to
develop an infinite slope at its end point. Further deformation curves
with larger $I$ all end similarly at a deformation of about $0.6$.  We
performed simulations for a number of different lattice sizes and the
exact value of $I_c$ depends on the cell dimensions, with increasing
wall separation causing a shift to smaller $I_c$ \cite{LE}.

As explained earlier, our ramp algorithm is designed to allow only a
certain amount of deformation for each new stationary state. So,
before the algorithm reaches a region of infinite slope it will force
increasingly smaller shear-rate increments; when these fall below the
pre-set threshold the algorithm will save the last configuration and
exit. To check the fate of these droplets we reloaded the last
configurations manually and again increased the shear rate by a very
small amount. All these droplets then broke up into two droplets; none
into three.

\begin{figure}
\resizebox{\columnwidth}{!}{\includegraphics{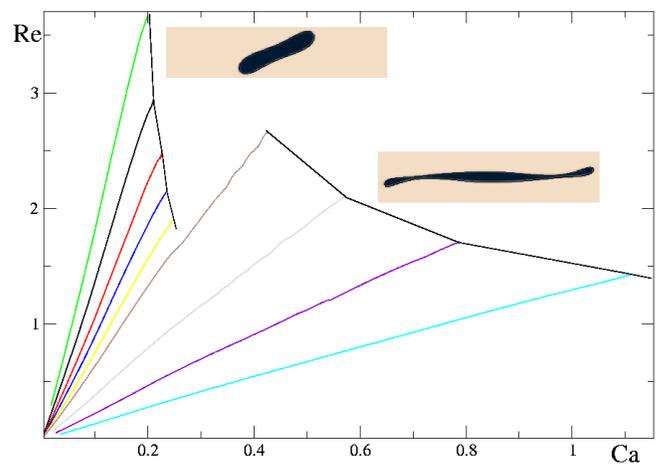}}\\
\caption{Deformation curves plotted in the Reynolds number / capillary
number plane. Note that the zero Re curve coincides with the x-axis
and does not end in breakup. The envelope of breakup points is shown
to guide the eye: this has two disconnected parts, corresponding to
two-lobe and three-lobe droplets at the last stable
configuration. Typical shapes are as indicated.}
\label{figCaRe}
\end{figure}

We show the dependence of the last stable capillary number on Re in
Figure \ref{figCaRe}.  Each of the runs appears an (almost) straight
line on the plot, ending at the breakup point. Two different critical
capillary number regimes are clearly separated in the Re/Ca plane.
Figure \ref{figCaRe} gives examples of the last stable droplet shapes
for both $I< I_c$ and $I>I_c$. For $I<I_c$ we find that the droplets
develop a double dimple; beyond the critical capillary number
Ca$_c(I)$ these break up into three. For $I > I_c$ the last stable
droplet shows one dimple; beyond Ca$_c$ these break up into only two
droplets.

Note that, in these simulations, we only observed breakup for Re $
<1.2$. The shape of the breakup envelope in Figure \ref{figCaRe},
together with the knowledge that the zero Reynolds number case does
not lead to break-up, suggests that the envelope could have a
horizontal asymptote at some nonzero value of Re (perhaps close to
unity). This would imply a finite critical Re below which no breakup
could occur, however large Ca. However, more detailed work would be
needed to reach a firm conclusion on this point.

\begin{figure}
\resizebox{\columnwidth}{!}{\includegraphics{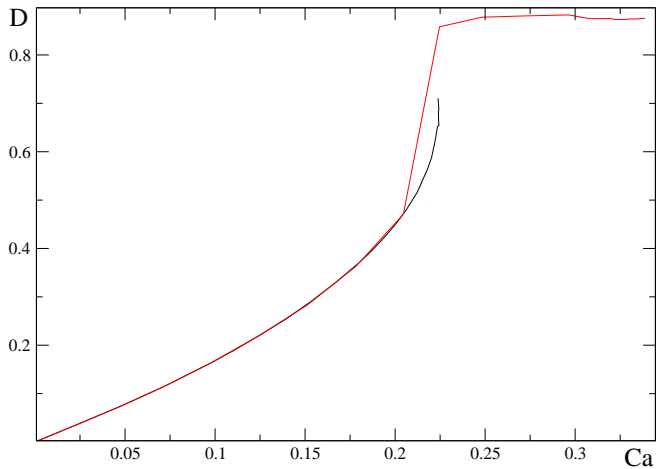}}\\
\caption{Two deformation curves for $I=11$. For one run with the
deformation restriction was removed. The droplet overshoots the
underlying break-up point and jumps into a three lobe structure, which
survives to larger capillary numbers until eventually breaking into
three droplets.}
\label{figDDef}
\end{figure}

We also performed simulations with a less sophisticated ramp algorithm
in which the shear rate was increased as before, but without
restriction on the deformation increment. In this case, different
behaviour was seen around $I_c$. Droplets which, when ramped more
carefully, would show the two-lobed breakup mode (with a vertical
tangent to the deformation curve at its end point; Figure \ref{figD})
could now jump past this onto the upper branch of the deformation
curve, developing instead a stable three-lobed shape, finally breaking
at higher Ca into three droplets. The deformation curve for one such
case is compared to the more careful ramp for the same $I$ in Figure
\ref{figDDef}. This proves that the details of the flow history can
strongly affect the final breakup mode and also the shear rate at
which it occurs.

Comparing this result with the general trend shown for larger $I$ in
Figure \ref{figD} suggests a conjecture: that the behaviour is
governed by an underlying, continuous deformation curve that has, for
$I>I_c$, folded over into an S-shape. With such a curve, one might
naturally expect a discontinuous jump in deformation on ramping up to
the point of vertical tangency. However, it appears that droplets
cannot survive the jump (but instead break in two) if brought very
close to this point before the jump is made. With a faster ramp, the
jump occurs sooner; the droplet then can survive the jump and continue
onto the upper branch of the deformation curve. The mechanism seems to
involve transient stretching just after the shear rate is incremented;
this allows the droplet to find the upper branch of the deformation
curve, but only if the shear rate increment is sufficient. Note that
such an S-shaped deformation curve would also admit hysteresis, in
which two states of deformation were both possible at exactly the same
Ca. However, we never got clear evidence of this in our simulations,
so any hysteresis loop appears to be relatively small.

{\em Conclusions:} In this article we have provided evidence that
droplet breakup in two dimensions requires a nonzero Reynolds
number. The dependence of the critical capillary number Ca$_c$ on the
dimensionless number $I$= Re/Ca for two-dimensional droplet break-up
shows an unexpected transition at $I=I_c$. (The precise value of $I_c$
depends on the wall separation.) This transition is consistent with an
evolution of the underlying deformation curve $D($Ca$)$ into an
S-shape. (However we did not observe hysteresis around this
transition.) A slow enough ramp leads to a two-lobed droplet breakup
mode at Ca$_c$ for $I > I_c$ whereas under different ramp conditions a
droplet can, before reaching Ca$_c$, jump from a two lobed structure
to a three lobed one, which is then stable to much higher Ca. For $I <
I_c$ the droplet evolves to a three lobed structure before breaking,
no matter how slow the ramp rate, and then breaks into three droplets.
From extrapolation of the critical capillary number we proposed that,
even at very large Ca, a threshold in Re of order unity may need to be
exceeded; but an alternative is that the Ca$_c$ diverges only as
Re$\to 0$.

Because of the care required in achieving equilibration close to
Ca$_c$ it would be difficult to perform similar numerical studies in
three dimensions without very substantial resources. Nonetheless, an
important open question is whether, on increasing the relative
importance of inertia, the breakup mode in three dimensions also
undergoes a sudden transition as we discovered here. We are unaware of
experiments addressing this, and note that in such experiments (as
well as in simulations) very careful control of the shear history
might be needed to identify any transitions that may be present.

{\em Acknowledgments:} Work funded in part by EPSRC Grants GR/M56234
and GR/R67699.

\def\jour#1#2#3#4{{#1} {\bf #2}, #3 (#4).}
\def\tbp#1{{\em #1}, to be published.}
\def\inpr#1{{\em #1}, in preparation.}
\def\tit#1#2#3#4#5{{#1} {\bf #2}, #3 (#4).}

\def\ap{Adv. Phys.}
\def\arf{Ann. Rev. Fluid Mech.}
\def\epl{Euro. Phys. Lett.}
\def\ijmp{Int. J. Mod. Phys. C}
\def\jcp{J. Chem. Phys.}
\def\jpc{J. Phys. C}
\def\jpcs{J. Phys. Chem. Solids}
\def\jpco{J. Phys. Cond. Mat}
\def\jsp{J. Stat. Phys.}
\def\jf{J. Fluids}
\def\jfm{J. Fluid Mech.}
\def\jnnfm{J. Non-Newtonian Fluid Mech.}
\def\pfa{Phys. Fluids A}
\def\prl{Phys. Rev. Lett.}
\def\pr{Phys. Rev.}
\def\pra{Phys. Rev. A}
\def\prb{Phys. Rev. B}
\def\pre{Phys. Rev. E}
\def\pa{Physica A}
\def\pla{Phys. Lett. A}
\def\ps{Physica Scripta}
\def\roy{Proc. Roy. Soc.}
\def\rmp{Rev. Mod. Phys.}
\def\zpb{Z. Phys. B}


\begin{thebibliography}{200}
 
\bibitem{taylor}
G.I.~Taylor,
\tit{Proc. Roy. Soc.}{26}{501}{1934}{}

\bibitem{renardy1}
Y.Y.~Renardy and V.~Cristini,
\tit{Phys. of Fluids}{13}{7}{2001}{}

\bibitem{renardy2}
Y.Y. Renardy and V. Cristini,
\tit{Phys. of Fluids}{13}{2161}{2001}{}

\bibitem{xi}
H. Xi and C. Duncan,
\tit{Phys. Rev. E}{59}{3022}{1999}{}

\bibitem{li}
J. Li, Y.Y. Renardy and M. Renardy,
\tit{Phys. of Fluids}{12}{269}{2000}{}

\bibitem{halliday}
I. Halliday, C.M. Care, S. Thompson and D. White,
\tit{\pre}{53}{1602}{1996}{}

\bibitem{sheth}
K.S. Sheth and C. Pozrikidis,
\tit{Computers and Fluids}{24}{101}{1995}{}

\bibitem{ludwig}
J.C. Desplat, I. Pagonabarraga, P. Bladon,
\tit{Computer Physics Communications} {134}{273}{2001}{}

\bibitem{kendon}
V.M. Kendon, M.E. Cates, I. Pagonabarraga, J.C. Desplat and P. Bladon,
\tit{J. Fluid Mech.} {440}{147}{2001}{}

\bibitem{foot}
In particular, it was found that unwanted diffusive currents could arise near the tips of sheared droplets if the diffusivity was too {\em small}; by increasing it, the currents were confined to a very thin region near the interface.

\bibitem{wagner}
A.J. Wagner and J.M. Yeomans,
\tit{\ijmp}{8}{773}{1997}{}

\bibitem{LE}
The maximal achievable shear in a lattice Boltzmann simulation is
limited by $\dot{\gamma}^{max} \simeq 0.2/L$ where $L$ is the distance
between the walls. This made it impossible for get a meaningful estimate of the
limit of $I_c$ for infinite $L/R_0$. A newly developed
Lees-Edwards condition for lattice Boltzmann may make it
possible to circumvent the restrictions for the shear-rate in the
future; see: A.J. Wagner and I. Pagonobarraga,
\tit{J. Stat. Phys.} {107}{521}{2002}{}


\end{thebibliography}
\end{document}